\documentclass [12pt]{article}
\usepackage{amsmath}
\usepackage{amsfonts}
\usepackage{epsfig}
\begin{document}
\def\b{\bar}
\def\d{\partial}
\def\D{\Delta}
\def\cD{{\cal D}}
\def\cK{{\cal K}}
\def\f{\varphi}
\def\g{\gamma}
\def\G{\Gamma}
\def\l{\lambda}
\def\L{\Lambda}
\def\M{{\Cal M}}
\def\m{\mu}
\def\n{\nu}
\def\p{\psi}
\def\q{\b q}
\def\r{\rho}
\def\t{\tau}
\def\x{\phi}
\def\X{\~\xi}
\def\~{\widetilde}
\def\h{\eta}
\def\bZ{\bar Z}
\def\cY{\bar Y}
\def\bY3{\bar Y_{,3}}
\def\Y3{Y_{,3}}
\def\z{\zeta}
\def\Z{{\b\zeta}}
\def\Y{{\bar Y}}
\def\cZ{{\bar Z}}
\def\`{\dot}
\def\be{\begin{equation}}
\def\ee{\end{equation}}
\def\bea{\begin{eqnarray}}
\def\eea{\end{eqnarray}}
\def\half{\frac{1}{2}}
\def\fn{\footnote}
\def\bh{black hole \ }
\def\cL{{\cal L}}
\def\cH{{\cal H}}
\def\cF{{\cal F}}
\def\cP{{\cal P}}
\def\cM{{\cal M}}
\def\ik{ik}
\def\mn{{\mu\nu}}
\def\a{\alpha}

\title{Conflict quantum theory and gravity as a source of particle stability}

\author{Alexander Burinskii \\
Theor.Physics Laboratory, NSI, Russian Academy of Sciences,\\ B. Tulskaya 52 Moscow 115191 Russia,
email: bur@ibrae.ac.ru}

\date{\emph{Essay written for the Gravity Research Foundation 2015\\
Awards for Essays on Gravitation}\\
March 31, 2015 } \maketitle

.
\begin{abstract}
We build a regular core of the Kerr-Newman $(KN)$ solution and considered  it as an extended
soliton or bag model of spinning particle creating external gravitational and electromagnetic field
of an electron. The known conflict between Quantum Theory and Gravity is solved by formation of a
domain wall boundary separating zones of their influence. Gravity controls external classical
$(KN)$ spacetime, while Quantum theory is responsible for the flat zone of supersymmetric core. We
show that Bogomolnyi bound determines precise boundary between these zones, controlling the shape,
dynamics and stability of the particle.
\end{abstract}


\newpage

 Modern physics is based on Quantum theory
and Gravity. The both theories are confirmed experimentally with
great precision. Nevertheless, they are conflicting and cannot be
unified in a whole theory.
The most acutely this conflict  manifests itself in the problem of
structure of elementary particles, which are pointlike in quantum
theory, but  should be extended  for consistency with gravity.
Extended particlelike models (solitons), such as Q-balls, bags, skirmions and vortex strings are
widely discussed now.

We will focus here on the famous $KN$ solution which was originally proposed as the external gravitational and
 electromagnetic field of a rotating charged body \cite{DKS}, but it turns out, in good agreement with the external field of the electron \cite{DKS,Car}. By nature, it is gravitating and, exhibiting relationships to electron, it can shed light on the causes of conflict and ways of removing it.

\noindent The Kerr-Schild form of metric \cite{DKS}
\be g_\mn =\eta_\mn + 2H k_\m k_\n , \quad H=\frac {mr -e^2/2}{r^2+a^2 \cos ^2 \theta}\label{KSH} \ee in
which $ \eta_\mn $ is metric of auxiliary Minkowski space $M^4 ,$ (signature $(- + + +)$), $H$ is a scalar function,  $r$ and $\theta$ are ellipsoidal coordinates and $ k_\m $ is a null vector
field, $ k_\m k^\m =0 ,$ forming a vortex polarization of Kerr space-time -- the Principal Null Congruence (PNC) $\cal K .$  The surface $r=0$ represents a disklike "door" from
negative sheet $r<0$ to positive one $r>0$. The smooth extension of the solution from retarded to advanced sheet (together with smooth extension of the Kerr PNC) occurs   via
disk $ r=0 $ spanned by the Kerr singular ring $ r=0, \ \cos\theta=0 $ (see fig.1).  The null vector fields $k^{\m\pm}(x)$ turns out to be different on these
sheets, and  two different null congruences ${\cal K}^\pm ,$ create two different metrics $
g_\mn^\pm =\eta_\mn + 2H k_\m^\pm k_\n^\pm $ on the same Minkowski background.

  The mysterious two-sheeted structure of the Kerr geometry caused
  searching diverse models for source of the KN solution avoiding negative sheet.
  Relevant "regularization" of this space was suggested by L\'opez \cite{Lop}, who excised singular region together with negative sheet and replace it by a regular core with a flat internal metric $\eta_\mn .$ It formed a \emph{vacuum bubble} which should be matched with external KN solution along the boundary $r=R,$ determined by the condition
  \begin{equation}
H|_{r=R}(r)=0 \quad \Rightarrow \quad R = r_e = \frac {e^2}{2m} . \label{Hre}
\end{equation}
 Since $r$ is the Kerr oblate ellipsoidal coordinate, see Fig.2, for $R>0 $  bubble covers the Kerr singular ring, forming a thin rotating disk of radius $r_c \sim a = \hbar/mc ,$  and oblateness of the disk $r_e/r_c$ corresponds to fine structure constant $r_e/r_c \sim e^2 = \alpha \sim 137^{-1}.$

\begin{figure}[ht]
\centerline{\epsfig{figure=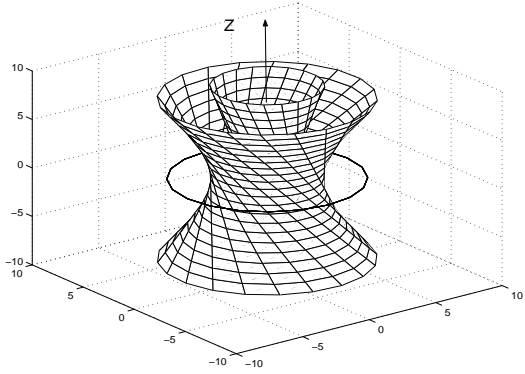,height=5cm,width=7cm}}
\caption{Vortex of the Kerr congruence. Twistor null lines are
focused on the Kerr singular ring, forming a circular
gravitational waveguide, or string with lightlike excitations.}
\end{figure}

\begin{figure}[ht]
\centerline{\epsfig{figure=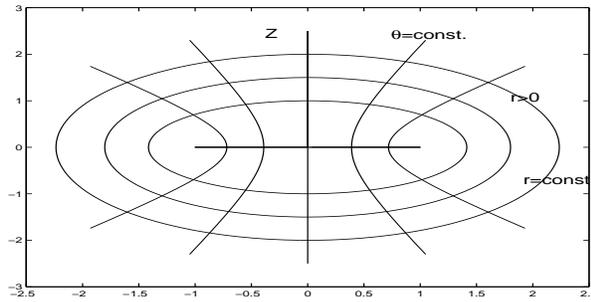,height=4cm,width=8cm}}
\caption{Kerr's oblate spheroidal coordinates  cover
space-time twice, for $r>0$ and $r<0$.}
\end{figure}

Development of the L\'opez bubble model  led to a soliton model of a vacuum bubble with domain wall
boundary, where Gravity controls external classical space-time, while the Quantum theory forms a
supersymmetric pseudo-vacuum state inside the soliton on the base of Higgs mechanism of breaking
symmetry. In this case conflict  between Quantum Theory and Gravity is avoided by
\textbf{principles of separation of zones of influence:}

\begin{description}
         \item[\textbf{PI:}] \textbf{\emph{space-time should be flat inside the core,}}
         \item[\textbf{PII:}]\textbf{ \emph{exterior should be exact KN solution}.}
         \item[\textbf{PIII:}]\textbf{ \emph{boundary between PI and PII is determined by the L\'opez condition (\ref{Hre})}.}
       \end{description}

In \cite{BurSol,BurSol1,Tomsk} was mentioned mysterious effectiveness of these principles, since they allow almost uniquely determine features of the KN core.

Now we see that boundary between these zones is precisely determined by \emph{Bogomolnyi bound}, and thus, the features,
dynamics and stability of the KN particle are determined by supersymmetry. But supersymmetry is a classical limit of quantum theory \cite{Rajar}, and thus, Bogomolnyi equations represent bosonic sector of quantum particle,  responsible for the domain wall phase transition and external gravity.
\bigskip

\bigskip

\noindent\textbf{Gravitating bag.} Bubble-source formed by L\'opez boundary was generalized to soliton \cite{BurSol,BurSol1})  or a bag \cite{BurEmerg,Tomsk,BurBag}. The bag model conception allows to incorporate fermionic sector in which mass of the Dirac equation is generated by Higgs field via Yukawa coupling \cite{SLAC}. Boundary of the bag is modelled by a domain-wall interpolating between  external KN solution and flat internal  pseudo-vacuum state, and  phase transition between these states is controlled by the Higgs mechanism of symmetry breaking. However, the used in typical soliton and bag models quartic potential for the Higgs field $\Phi ,$ \be V(|\Phi|)=g(\bar\sigma \sigma - \eta^2)^2 , \ \  \sigma =<|\Phi|> , \label{phi4} \ee is not suit since it distorts external electromagnetic field of  KN solution.

  Contrary to the standard bag model forming  \emph{a cavity in the Higgs condensate}, \cite{MIT}, the condition $\textbf{PII}$ requires the Higgs condensate be enclosed \emph{inside the bag.} This cannot be done with potential (\ref{phi4}), and $\textbf{PII}$  requires the use of complex supersymmetric scheme of the phase transition, described by three chiral fields
$\Phi^{(i)}, \ i=1,2,3 .$  One of the fields, say $\Phi^{(1)} ,$ is identified as the Higgs field $\Phi ,$  and  we set new notations \be (\Phi, Z, \Sigma) \equiv (\Phi^1, \Phi^2, \Phi^3) . \ee

Due to condition $\textbf{PI} , $ bag is placed in flat region, and domain
wall phase transition may be considered with flat metric, $g_\mn = \eta_\mn .$ Therefore the domain
wall boundary of the bag and the bag as a whole are nor dragged by rotation. Because of that, the chiral part of the
Hamiltonian  is simplified to \be H^{(ch)} = T_0^{ ~0  (ch)} = \frac 12
\sum_{i=1}^3 ~ [ \sum_{\m=0}^3  |\cD^{(i)}_\m \Phi^i|^2 +  |\d_i W|^2 ], \label{HamCh} \ee where the
covariant derivatives $\cD^{(i)}_\m \equiv \d_i -ie A^i_\m$ are flat. Following \cite{WesBag}, the potential $V$ is expressed via superpotential $ V(r)=  \sum _i |\d_i W|^2 ,$ and the condition  $\d_i W =0 $ determines two vacuum states:

(I) external: $r>R +\delta $, $V (r) = 0 , \ \Phi =0 $,   and

(II) internal: $r<R-\delta $, $V (r) = 0 , \ |\Phi|
= \eta = const. $

\noindent These vacua are separated by a positive spike of potential $V$ in the flat transition zone $R-\delta < r <R-\delta ,$ see Fig.3.

\begin{figure}[ht]
\centerline{\epsfig{figure=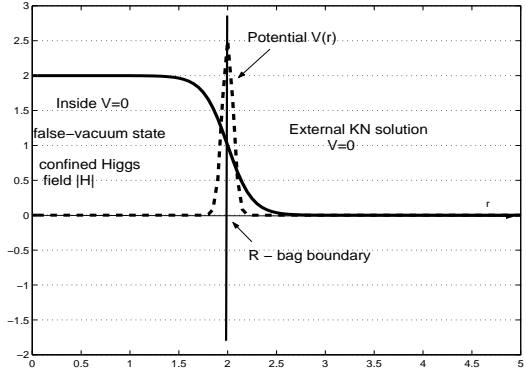,height=5cm,width=7cm}}
\caption{Symmetry breaking in the KN soliton-bag model.
Potential $V(R)$ forms the inner and outer vacuum states $V=0$
with a narrow spike at edge of source.}
\end{figure}

Our task will be to find Bogomolnyi equation and bound which determine mass-energy carried by the soliton-bag.
Stationarity of the problem and axial symmetry allow us to separate the terms related
with $\phi$ and $t$- dependence from radial evolution of the spheroidal domain wall surface, Fig.4.

\noindent Search soliton solutions requires sometimes guesses for a special conversion, \cite{Rajar}.
The phase of complex chiral fields represents the problem of this sort for Bogomolnyi equations, and a special trick
was considered for this case in \cite{Cvet91,GibTown}.

\begin{figure}[ht]
\centerline{\epsfig{figure=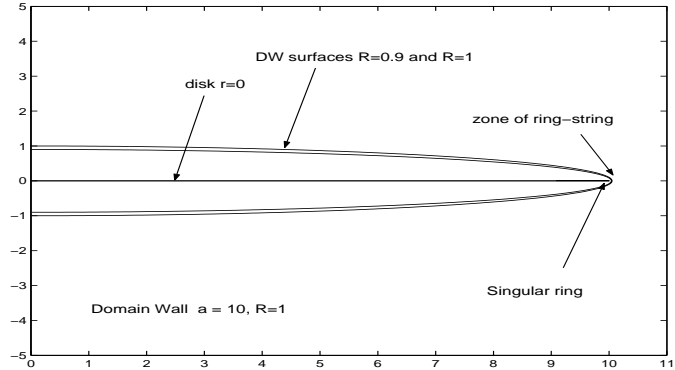,height=5cm,width=9cm}}
\caption{Axial section of the spheroidal domain wall phase transition.}
\end{figure}

Consider first internal zone (II), where amplitudes of all chiral fields are constant.

We obtain that  inside the bag survive only two quadratic terms in (\ref{HamCh})
\be T_0^{ ~0  (ch-in)} = \frac 12 [ |\cD^{(1)}_0 \Phi^1|^2 + |\cD^{(1)}_\phi \Phi^1|^2], \label{MChIn} \ee
and Bogomolnyi equations for minimum of Hamiltonian are reduced  here to simple equations
\be (\d_0 -ie A_0^{in}) \Phi^1=0, \quad (\d_\phi - ie A_\phi^{in}) \Phi^1=0 , \label{Aphi0}\ee
where $A_0^{in}$ and $A_\phi^{in}$ are components of the KN vector potential
\be A_\m
dx^\m = - Re \ [(\frac e {r+ia \cos \theta})] (dr - dt - a \sin ^2 \theta d\phi ) \label{Am} \ee
inside the bag.
  \begin{figure}[h]
\centerline{\epsfig{figure=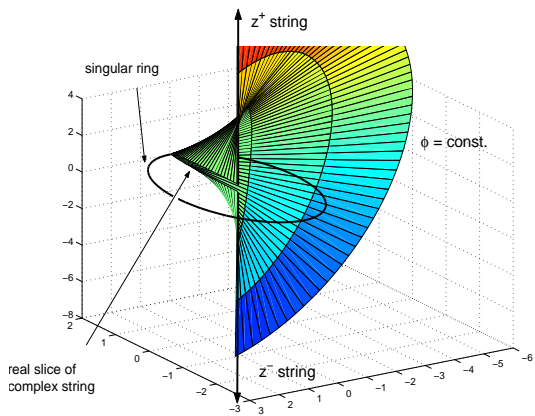,height=5.5cm,width=7cm}}
\caption{\label{label} Kerr's coordinate $\phi=const.$   Vector potential, dragged by singular ring, forms closed loop
along edge border of bag.}
\end{figure}

We note that although metric is flat, \emph{influence of gravity is saved in the
shape of the bag and also in dragging of the vector potential}, which
take simple form only in the Kerr oblate coordinate system, and we adopt this system also for  calculations in transition zone.
From (\ref{MChIn}) and (\ref{Am})  we obtain \be \Phi^1 = |\Phi| e^{i\chi(t,\phi)}, \quad  \chi(t,\phi)=2\omega t + \phi , \label{Phi1}\ee which yields two important results, obtained earlier in \cite{BurSol,BurSol1}:

\textbf{(A)} the Higgs field \emph{oscillates} with the frequency $\omega= 2m $  -- KN bag is oscillon \cite{oscil},

 \textbf{(B)} the component $A_\phi$ forms closed loop along boundary of the bag
 in equatorial plane, see Fig.5, which leads to
 \emph{quantization of the angular momentum}  $ J=n/2,
\quad n=1,2,3,... \label{Jn/2} .$

\bigskip
The problem is now reduced to a single radial coordinate, where we use the trick suggested in \cite{Cvet91,GibTown} for much more simple case of one chiral field and planar domain wall.  We rewrite the mass-energy of the chiral bubble in the form of Bogomolnyi integral
 over one variable, which is the Kerr oblate radial coordinate $r.$

\noindent   After ansatz \be \Phi (x) \equiv \Phi^1(x) = |\Phi^1 (r)| e^{i\chi(t,\phi)}, \label{Phichi}\ee
\be \Phi^2 = \Phi^2 (r), \quad \Phi^3 = \Phi^3 (r) , \ee all the fields depend only on the oblate spheroidal coordinate $r ,$ playing now the role similar to `parallel shift' for the planar domain walls,  see Fig.4.
 Using transformation suggested in \cite{Cvet91,GibTown} we can represent now (\ref{HamCh}) in the form
\be H^{(ch-r)} = \sum_{i=1}^3 \frac 12 | \d_r \Phi^i - \d W /\d
 \Phi^i |^2 + 2 Re \ (\d  W /\d  \Phi^i) \d_r \Phi^i .
\label{HamChd} \ee
From which we obtain the first order Bogomolnyi equations
 \be \cD^{(i)}_r \Phi^i = \d  W /\d \Phi^i , \quad \cD^{(i)}_r \bar\Phi^i = \d \bar W /\d \bar
\Phi^i , \label{Bog1}\ee
and the Bogomolnyi bound minimizing energy density $H^{(ch-r)} .$

It shows that the KN bag forms a topologically stable BPS-saturated configuration.
Using axial symmetry of the system, we can express the total mass in the Kerr oblate coordinates as follows

\be M_{ch}= 4\pi  \int r^2  \d_r W (r) dr \approx \frac 43 \pi  r_e^3 W(0) + 4\pi r_e^2 \Delta W ,
\label{MbubKerr}\ee where $\Delta W$ is bounce of the superpotential, crossing the domain wall.

\begin{figure}[ht]
\centerline{\epsfig{figure=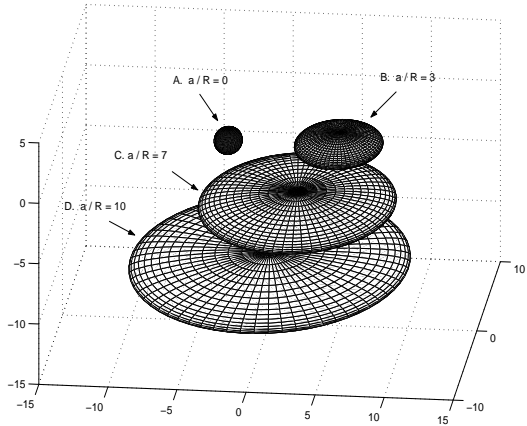,height=6cm,width=7cm}}
\caption{(A): Nonrotating spherical bag, $a/R =0$, and the rotating bags with (B): $a/R =3$;(C): $a/R =7$, (D):$a/R =10$.}
\label{fig3}
\end{figure}

A deviation of the domain wall boundary from this state increases energy of the bag, and therefore, stability of the
system is really based on stated in \textbf{PI-PIII} requirements  to  separate of the zones of influence for gravity and quantum theory.

 The typical  bags are flexible and take a stringlike form at excitations\cite{SLAC,Giles}. Similar features exhibits the KN bubble boundary (\ref{Hre}) under action of rotation, see Fig.6. Electromagnetic excitations lead to formation of the ring-string on the sharp border of the bag \cite{BurEmerg,BurBag,Tomsk}, and the requirements \textbf{PI-PIII} result in emergence of a singular pole associated with traveling-wave excitations of the KN ring-string \cite{Bur0,IvBur,BurSen,Tomsk}. This pole can be associated with a single quark, and finally, the KN bag takes the form of a coherent ``bag-string-quark'' system \cite{BurBag,Tomsk}.

\end{document}